\documentclass[epj,final]{svjour}
\journalname{EPJ B}
\usepackage[dvipdfm]{graphicx}
\usepackage{amstext}
\usepackage[usenames,dvipsnames]{color}
\def\pf6{(TM\-TSF)$_2$\-PF$_6$}

\begin{document}

\title{Anisotropy and field-dependence of the spin-density-wave dynamics in the quasi one-dimensional conductor (TMTSF)$_2$PF$_6$}


\author{P.~Zornoza \inst{1} \and K.~Petukhov \inst{1} \and M.~Dressel
\inst{1} \and N.~Biskup \inst{2} \and T.~Vuletic \inst{2} \and S.~Tomic
\inst{2}}

\institute{1.~Physikalisches Institut, Universit\"at Stuttgart,
Pfaffenwaldring 57, D-70550 Stuttgart, Germany \and Institut za fiziku,
P.O. Box 304, HR-10001 Zagreb, Croatia}

\date{Received: \today / Revised version: }

\abstract{ The anisotropic and non-linear transport properties of the
quasi one-dimensional organic conductor \pf6\ have been studied by dc,
radiofrequency, and microwave methods. Microwave experiments along all
three axes reveal that collective transport, which is considered to be
the fingerprint of the spin-density-wave condensate, also occurs in the
perpendicular $b^{\prime}$ direction. The pinned mode resonance is
present in the $a$ and $b^{\prime}$-axes response, but not along the
least conducting $c^*$ direction. The ac-field threshold, above which
the spin-density-wave response is non-linear, strongly decreases as the
temperature drops below 4~K. With increasing strength of the microwave
electric field and of the radiofrequency signal, the pinned mode and
the screened phason loss-peak shift to lower frequencies. In the
non-linear regime, in addition to the phason relaxation mode with
Arrhenius-like resistive decay, an additional mode with very long and
temperature-independent relaxation time appears below 4~K. We attribute
the new process to short-wavelength spin-density-wave excitations
associated with discommensurations in a random commensurate $N=4$
domain structure.
 }

\titlerunning{Anisotropic and field-dependent pinning of SDW}

\maketitle

\section{Introduction}
The spin-density-wave (SDW) ground state of one-dimen\-sional
conductors attracts considerable attention even after two decades of
intensive research. Most experimental studies have been performed on
the quasi one-dimensional organic Bechgaard salt
di-(tetra\-methyl\-tetra\-selena\-fulvalene)-hexa\-flouro\-phosphate,
denoted as \pf6, which soon became the model compound of this
phenomenon. Since a large number of results have been accumulated over
the years, the SDW state of \pf6\ can be considered as understood to a
fair extent. Nevertheless, there are certain issues which still wait
for a clarification \cite{Jerome82,Gruner94a,Gruner94b}.

At $T_{\rm SDW}=12$~K, \pf6\ undergoes a metal-insulator transition
below which a thermally activated transport is observed. The insulating
SDW ground state is a consequence of the instability of the quasi
one-dimen\-sion\-al Fermi surface. In particular NMR experiments reveal
that the nesting vector in (TMTSF)$_2$PF$_6$ is incommensurate: ${\bf
Q} = [0.5a^*, (0.24\pm 0.03) b^*, (-0.06\pm 0.20) c^*]$
\cite{Takahashi86b}.

\begin{figure}
\centering\resizebox{0.44\textwidth}{!}{\includegraphics*{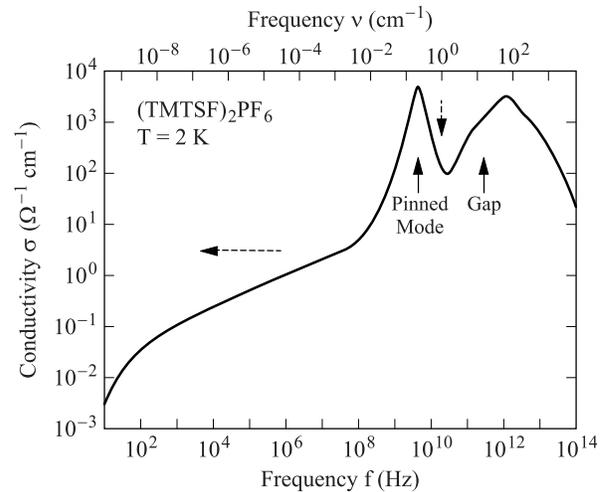}}
\caption{ \label{fig:10} Sketch of the frequency dependent SDW
conductivity composed using various sets of data along the chain axis
of \pf6. The solid arrows indicate the position of single particle gap
and the pinned mode resonance in the microwave frequency range (after
\protect\cite{Donovan94}). The dashed vertical and horizontal arrows
depict the frequency range of the investigations presented in this
work.}
\end{figure}

In the SDW state  the optical conductivity develops an absorption
edge in the infrared spectral range due to the opening of the
single-particle gap at the Fermi energy. While it turned out to be
extremely difficult to observe the gap in the direction along the
chains because the spectral range is strongly reduced in this
range of frequency (so-called clean limit), optical experiments
along the perpendicular direction \cite{Degiorgi96} unambiguously
proved the development of the SDW gap at
$2\Delta/(hc)=70$~cm$^{-1}$. The size of the SDW gap detected in
dc-limit is much smaller (approximately 30~cm$^{-1}$), probably
due to imperfections inducing in-gap states or due to the
dispersion of the energy bands in quasi one-dimensional compounds
\cite{Mihaly97}. A so-called pinned mode resonance is found around
5~GHz; it can be attributed to the collective response of the
condensate pinned to lattice imperfections \cite{Donovan94}. It
contains only a very small fraction of the spectral weight missing
after the collapse of the Drude response upon entering the
insulating SDW state. At even lower frequencies (in the range of
MHz, kHz and even below, depending on temperature) internal
deformations and screening by the conduction electrons lead to a
broad increase of the conductivity and corresponding broad
relaxational behavior. The overall behavior of the optical
response is plotted in Fig.~\ref{fig:10}.

A large number of investigations have been dedicated to the
non-linear transport of the SDW ground state
\cite{Donovan94,Tomic89,Mihaly91}. In very low fields $(E\ll
E_T$), as expected for a typical semiconductor, Ohmic behavior is
observed along the chains. However in electric fields which exceed
a characteristic threshold field $E_T$, a contribution to the
conductivity due to the sliding of the SDW was expected and
experimentally found. $E_T$ is typically in the 3-8~mV/cm range in
nominally pure specimens, with a factor of two increase of the
conductivity in fields which are several times larger than $E_T$.
The existence of both narrow-band noise \cite{Nomura89,Kriza91}
and Shapiro interferences \cite{Kriza91} above $E_T$ provide very
strong indirect evidence that the nonlinearity observed is
associated with a depinning of the SDW, however the large motional
narrowing of the proton lineshape seen above $E_T$ \cite{Clark93}
is by far the most conclusive proof. As expected for a SDW which
is pinned to impurities, the size of $E_T$ has been correlated to
the concentration of lattice defects induced by X-ray irradiation
\cite{Tomic89} as well as to the random disorder introduced by the
alloying at the anion sites.

In this Communication the issue of the anisotropic and field dependent
response of the collective mode is addressed. To our knowledge, the
non-linear conduction and the pinned mode resonance --~fingerprints of
the collective response~-- have only been investigated along the chain
direction. Also not much is known about the radiofrequency and
microwave response with increasing amplitude of the driving ac field.
Finally, the internal structure of the SDW and the origin of anomalies
observed in the spin-lattice relaxation rate, specific heat and
magneto-resistance, have not been consistently clarified yet
\cite{Takahashi87,Odin94,Basletic02}.

\section{Experimental Details}
Single crystals of the Bechgaard salt \pf6\ were grown by
electrochemical methods as described in Ref.~\cite{Dressel05}.
After several months we were able to harvest needle-shaped to
flake-like single crystals of several millimeters in length ($a$
axis) and a considerable width ($b^{\prime}$-direction) up to
2~mm. Here $a$ indicates the stacking direction of TMTSF
mole\-cules. Due the triclinic symmetry, $b^{\prime}$ denotes the
projection of the $b$ axis perpendicular to $a$, and $c^*$ is
normal to the $ab$ plane. The crystals were characterized by dc
resistivity measurements. Along the $a$-axis the experiments were
performed on needle-shaped samples with a typical dimension of
$2~{\rm mm}\times 0.5~{\rm mm}\times 0.1~{\rm mm}$ along the $a$,
$b^{\prime}$, and $c^*$ axes, respectively. The results on the
$b^{\prime}$-axis conductivity were obtained on a narrow slice cut
from a thick crystal perpendicular to the needle axis; the typical
dimensions of so-made samples were
$a~\times~b^{\prime}~\times~c^{*} = 0.2~{\rm mm}\times 1.3~{\rm
mm} \times 0.3~{\rm mm}$. Due to our advances in achieving large
sample geometry, we were able to measure $b^{\prime}$-axis
resistivity for the first time with basically no influence of the
$a$ and $c^*$ contributions and using standard four-probe
technique to eliminate the contact resistance. Also for the
$c^*$-axis transport, we were able to apply four contacts, two on
each side of the crystal. The contacts were made by evaporating
gold pads on the crystal, then 25~$\mu$m gold wires were pasted on
each pad with a small amount of silver or carbon paint. The \pf6\
samples were slowly cooled down to avoid cracks and ensure a
thermal equilibrium.

The real and imaginary parts of the conductance, $G(\omega)$ and
$B(\omega)$, were measured employing a HP4284A impedance analyzer
($f=20$~Hz to 1~MHz) in the two-probe configuration. Three samples with
lengths from 0.21 to 0.34~cm and cross-sections in the range of
$2.6\cdot 10^{-5}$ to $10^{-4}$~cm$^2$ were studied in the temperature
range between 1.5~K and 5~K; they all exhibited qualitatively the same
behavior. The data were taken by sweeping the frequency at a fixed
temperature. Both cooling and warming cycles were conducted and no
hysteresis was observed. In order to study the non-linear transport, an
ac signal amplitude $V_S$ was applied in the range between 0.07 and 1
of the dc threshold voltage $V_T$ for non-linear conductivity; which at
$T = 4.2$~K was in the range from $V_T=2$~mV to 4~mV. As the smallest
ac voltage delivered by the HP4284A is 5~mV, a home-made electronic
circuit was utilized to reduce the voltage effective on the sample to
the desired level. Dielectric functions were extracted from the
conductivity using the relations
$\varepsilon'(\omega)=B(\omega)/\omega$ and
$\varepsilon''(\omega)=[G(\omega)-G_0]/\omega$.  The observed
dielectric response can be well fitted by the phenomenological
Havriliak-Negami (HN) function widely used to describe relaxation
processes in disordered systems
\begin{equation}
\varepsilon(\omega) - \varepsilon_{\text{HF}} = \frac{\Delta
\varepsilon}{1+(i \omega \tau_{0})^{1-\alpha}} \label{HN}
\end{equation}
Here $\Delta\varepsilon = \varepsilon_0 - \varepsilon_{\text{HF}}$
is the relaxation strength, and $\varepsilon_0$ and
$\varepsilon_{\text{HF}}$ are the static and the high frequency
dielectric constant, respectively; $\tau_0$ denotes the mean
relaxation time and $1-\alpha$ is the shape parameter which
describes the symmetric broadening of the relaxation time
distribution function. More details on low-frequency dielectric
spectroscopy and the data analysis can be found in
Ref.~\cite{Pinteric01}.

In order to measure the microwave conductivity the crystals were placed
onto a quartz substrate and positioned in the maximum of the electric
field of different cylindrical copper cavities, resonating in the
TE$_{011}$ mode at 24, 33.5~GHz and 60~GHz, respectively. Along the
$a$-direction naturally grown needles could be used (typical dimensions
of $1~{\rm mm}\times 0.2~{\rm mm} \times 0.2~{\rm mm}$) since this
geometry is best for precise microwave measurements. To measure in
$b^{\prime}$ direction, we cut a slice
($a~\times~b^{\prime}~\times~c^{*} = 0.2~{\rm mm}\times 1.2~{\rm mm}
\times 0.2~{\rm mm}$) from a thick single crystal. In order to perform
micro\-wave experiments along the $c^*$ axis, a crystal was chopped
into several pieces (approximately cubes of 0.2~mm corner size) and
arranged up to four as a mosaic in such a way that a needle-shaped
sample of about $(0.2\times 0.2 \times 0.8)~{\rm mm}^3$ was obtained.
By recording the center frequency and the halfwidth of the resonance
curve as a function of temperature ($1.5~{\rm K}<T<300$~K) and
comparing them to the corresponding parameters of an empty cavity, the
complex electrodynamic properties of the sample, like the conductivity
and the dielectric constant, can be determined via cavity perturbation
theory; further details on microwave measurements and the data analysis
are summarized in \cite{Dressel05,Klein93,DresselGruner02}. Using  a
microwave attenuator in the waveguide line between Gunn diode and the
cavity allows us to reduce the power by -18~dB.

\section{Results}
\begin{figure}
\centerline{\includegraphics[width=8cm]{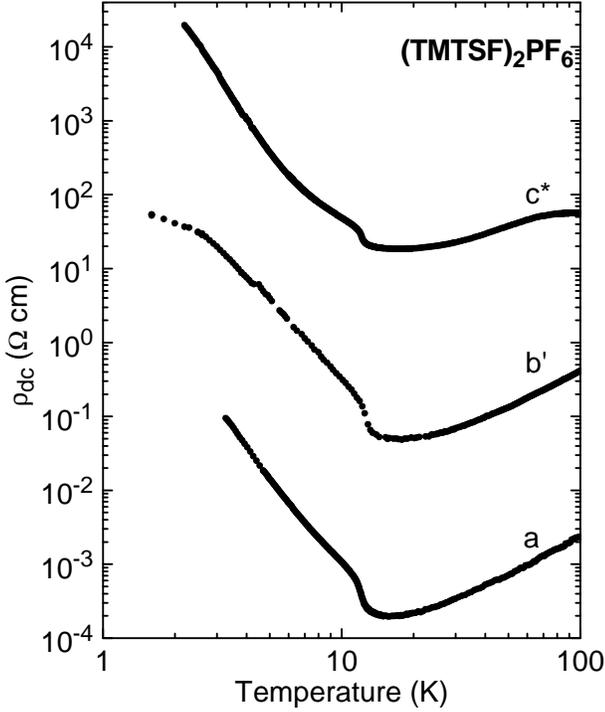}}
\caption{\label{fig:1}Temperature dependence of the dc resistivity of
(TMTSF)$_2$PF$_6$ single crystals along the $a$, $b^{\prime}$, and
$c^*$ crystallographic axes.}
\end{figure}

The temperature dependence of the dc resistivity is displayed in
Fig.~\ref{fig:1}, in double logarithmic representation where the
SDW transition at $T_{\rm SDW}=12$~K is clearly seen in all three
directions, $a$, $b^{\prime}$, and $c^*$. This is well understood
since in this compound the single-particle gap opens over the
entire Fermi surface. At lower temperatures the resistivity
increases steadily with some exponential temperature dependence
$\rho(T)\propto \exp\{\Delta/T\}$ due to excitations across the
energy gap ranging from $\Delta \approx 20$~K to 27~K
\cite{Petukhov04}. The results are in excellent agreement with the
predictions by the mean-field theory ($\Delta_0=1.76\,k_B T_{\rm
SDW}=21$~K) and previous findings \cite{Gruner94b}.

\begin{figure}
\centerline{\includegraphics[width=8cm]{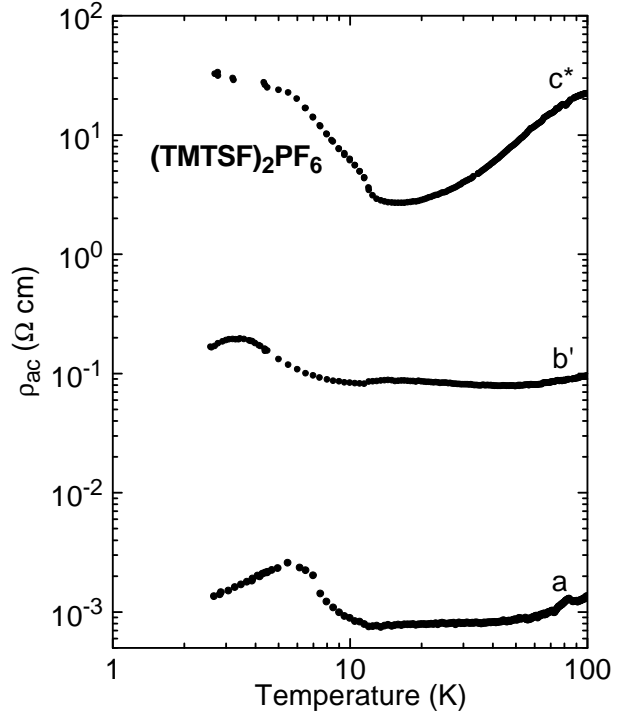}}
\caption{\label{fig:2}Temperature dependence of the microwave
resistivity of (TMTSF)$_2$PF$_6$ along the $a$, $b^{\prime}$, and $c^*$
crystallographic axes, measured at 33.5~GHz.}
\end{figure}

The results obtained for the microwave response
at 33.5~GHz are plotted in Fig.~\ref{fig:2}.
The SDW transition at $T_{\rm SDW} = 12$~K is again present in all
orientations. In contrast to the dc response, however,
for lowering the temperature the microwave resistivity
increases monotonously below $T_{\rm SDW}$ only in the $c^*$ direction.
For the electric field parallel to the chains and along the
$b^{\prime}$ direction, the behavior is more complicated;
the resistivity saturates below approximately 4~K
and in some cases even a maximum in $\rho_{ac}(T)$
is reached around $T\approx 4$~K
below which the resistivity decreases again.

\subsection{Anisotropic Response of the SDW Condensate}
\begin{figure}[h]
\centerline{\includegraphics[width=7.5cm]{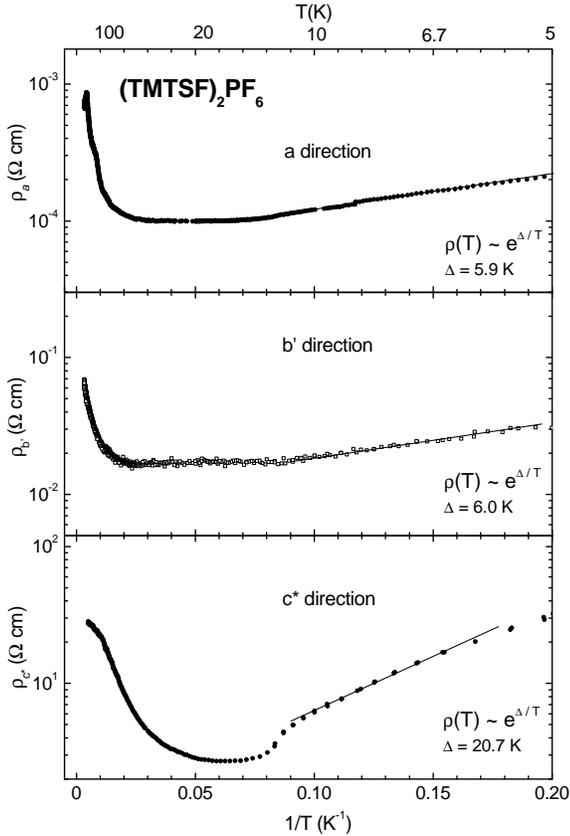}}
\caption{\label{fig:3}Arrhenius plot of the microwave resistivity of
\pf6\ along the $a$, $b^{\prime}$ and $c^*$ directions measured at
33.5~GHz.}
\end{figure}

In the temperature range slightly below $T_{\rm SDW}$, but still for
$T>4$~K, $\rho(T)$ may be described by an activated behavior, as
depicted in Fig.~\ref{fig:3} \cite{remark2}. The activation energy
along the $a$ and $b^{\prime}$ axes ($\Delta\approx 5.9$~K and 6.0~K,
respectively) is much smaller compared to the dc behavior, while for
the $c^*$ orientation the results at microwave frequencies perfectly
agree with the dc profile: $\Delta=20.7$~K.

\subsection{Field Dependent Response  of the SDW Condensate}
\begin{figure}
\centerline{\includegraphics[width=8cm]{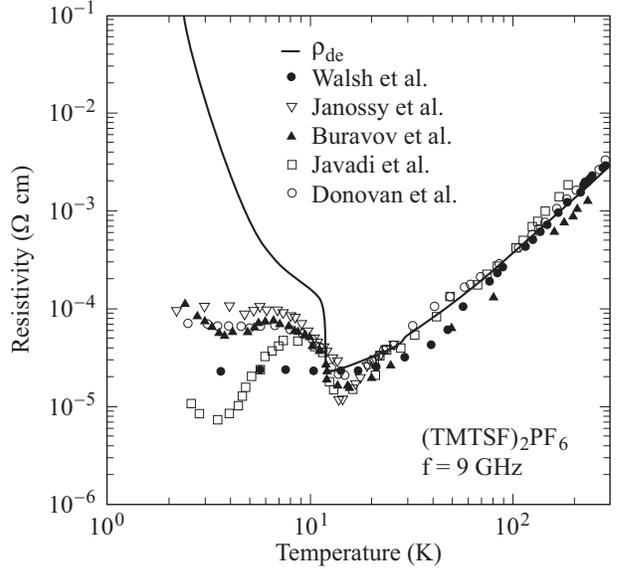}}
\caption{\label{fig:4}Comparison of the temperature dependent microwave
resistivity of (TMTSF)$_2$PF$_6$ along the $a$ axis reported by
different groups. Data taken from Walsh et al. \protect\cite{Walsh80},
Janossy et al. \protect\cite{Janossy83}, Buravov et al.
\protect\cite{Buravov85}, Javadi et al. \protect\cite{Javadi85}, and
Donovan et al. \protect\cite{Donovan94}. For better comparison the data
have been normalized to the 100~K dc value.}
\end{figure}

Even more surprising is the decreasing resistivity for temperatures
$T<4$~K.
Although the actual shape, its temperature dependence and magnitude
varies from sample to sample, the non-monotonous temperature dependence
and overall behavior is robust.
Up to six samples of different batches have been studied for each
orientation leading to very similar findings.
In fact, this puzzling behavior
was occasionally reported by different groups
\cite{Donovan94,Walsh80,Janossy83,Buravov85,Javadi85} over the years,
unfortunately without providing any meaningful explanation.
In Fig.~\ref{fig:4} the results for measurements conducted at
approximately 9~GHz are compiled.

Donovan {\it et al.} \cite{Donovan94} also investigated
the microwave response at different frequencies from 3 to 150~GHz,
and found that the deviations from the dc transport become less
when going to higher-frequency resonance cavities \cite{remark1}.
Note, however, that the output power of the microwave source and thus
the electric field strength inside the cavity strongly decreases with
higher frequencies.
It is also interesting to note,
that the effect is less pronounced in those cases where the sample is
located in the maximum of the magnetic field
compared to measurements in the electric field maximum of the cavities.
The present work tries to disentangle the
influence of frequency and field strength.

\begin{figure}
\centerline{\includegraphics[width=8cm]{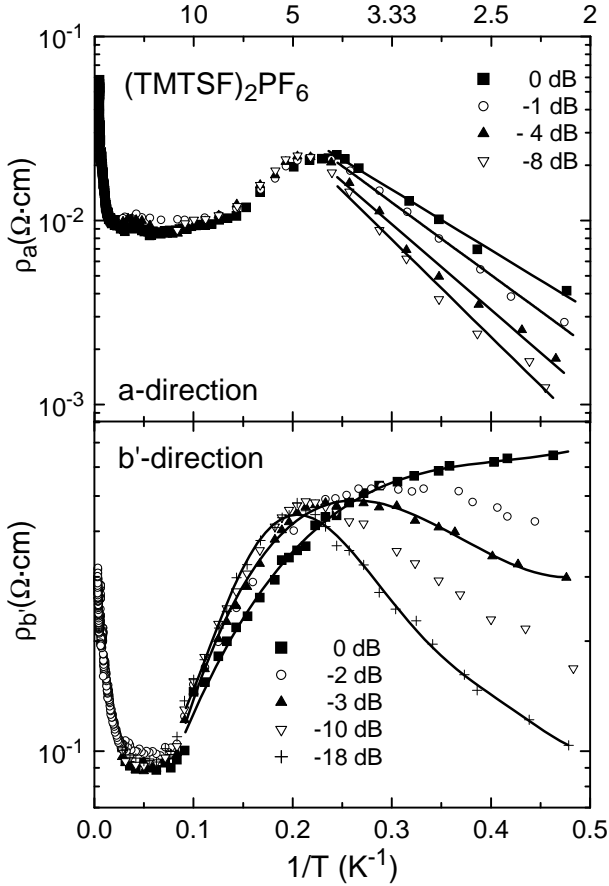}}
\caption{\label{fig:5}Resistivity of (TMTSF)$_2$PF$_6$ as a function of
inverse temperature (Arrhenius representation) for different
attenuation of the applied microwave power as indicated, leading to
different strength of the microwave electric field inside the cavity.
The experiments are performed at 33.5 GHz with the electric field
oriented (a) parallel and (b) perpendicular to the chain direction.}
\end{figure}

In order to further explore the non-monotonous behavior of
$\rho_{ac}(T)$, we performed the cavity perturbation experiments
along the $a$ and $b^{\prime}$ orientation using different
microwave power, i.e.\ as a function of the electric field
strength inside the cavity. The output power of the Gunn
oscillator was attenuated from 0 to -18~dB. The power $P_{\rm
input}$ coupled into the cavity varies from $120~\mu$W to
$2.5~\mu$W, resulting in a reduction of the average electric field
$E\propto\sqrt{P_{\rm input}}$ from 54~mV/cm to 8~mV/cm; in the
antinode the field strength is larger by a factor of two. Note,
however, that the absolute values only give a rough estime (within
a factor of 5) due to uncertainties in the coupling to the cavity
and losses in the waveguides. While Walsh et al. \cite{Walsh80}
estimate 1~mV/cm in their cavity, Buravov et al. \cite{Buravov85}
give 30~mV/cm for their 9~GHz measurement. The latter group
observed no change of the response when the electric field is
reduced by a factor of 4. In Fig.~\ref{fig:5} the microwave
resistivity is plotted as a function of inverse temperature for
different values of the microwave power \cite{remark2}. Reducing
the power leads to a sharper increase below $T_{\rm SDW}$ and the
maximum in $\rho(T)$ shifts to higher temperature; the drop of the
low-temperature resistivity becomes even more pronounced. This
behavior is robust and was confirmed for different samples.
Although  this sort of field dependence is more or less observed
for both orientations, along the $a$ and $b^{\prime}$ directions,
due to the higher resistivity, the experiments are more precise
for $E\parallel b^{\prime}$ and can cover a larger range of
applied microwave power.

\begin{figure}
\centerline{\includegraphics[width=7cm]{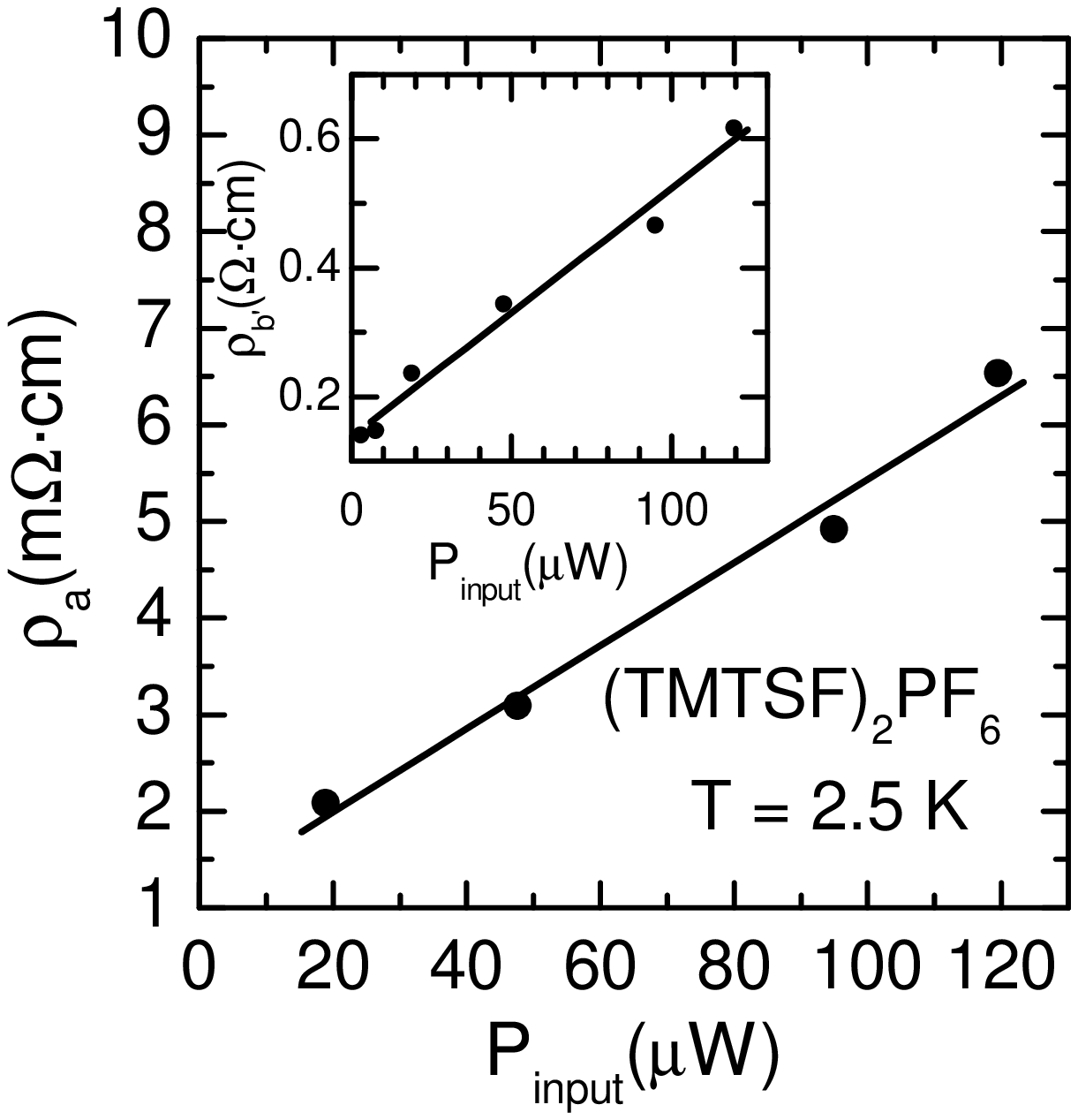}}
\caption{\label{fig:6}Resistivity of \pf6\ along the chain direction of
\pf6\ measured at $T=2.5$~K versus power put into the 33~GHz cavity.
The inset shows the power dependence in the $b^{\prime}$ direction.}
\end{figure}
In Fig.~\ref{fig:6} the power dependence of the resistivity $\rho_a$
and $\rho_{b^{\prime}}$ is plotted for a fixed temperature $T=2.5$~K.
In both cases a linear dependence on the input power is found with an
increase by a factor of 4 in resistivity when the input power changes from
3~$\mu$W above 120~$\mu$W.

\begin{figure}
\centerline{\includegraphics[width=6cm]{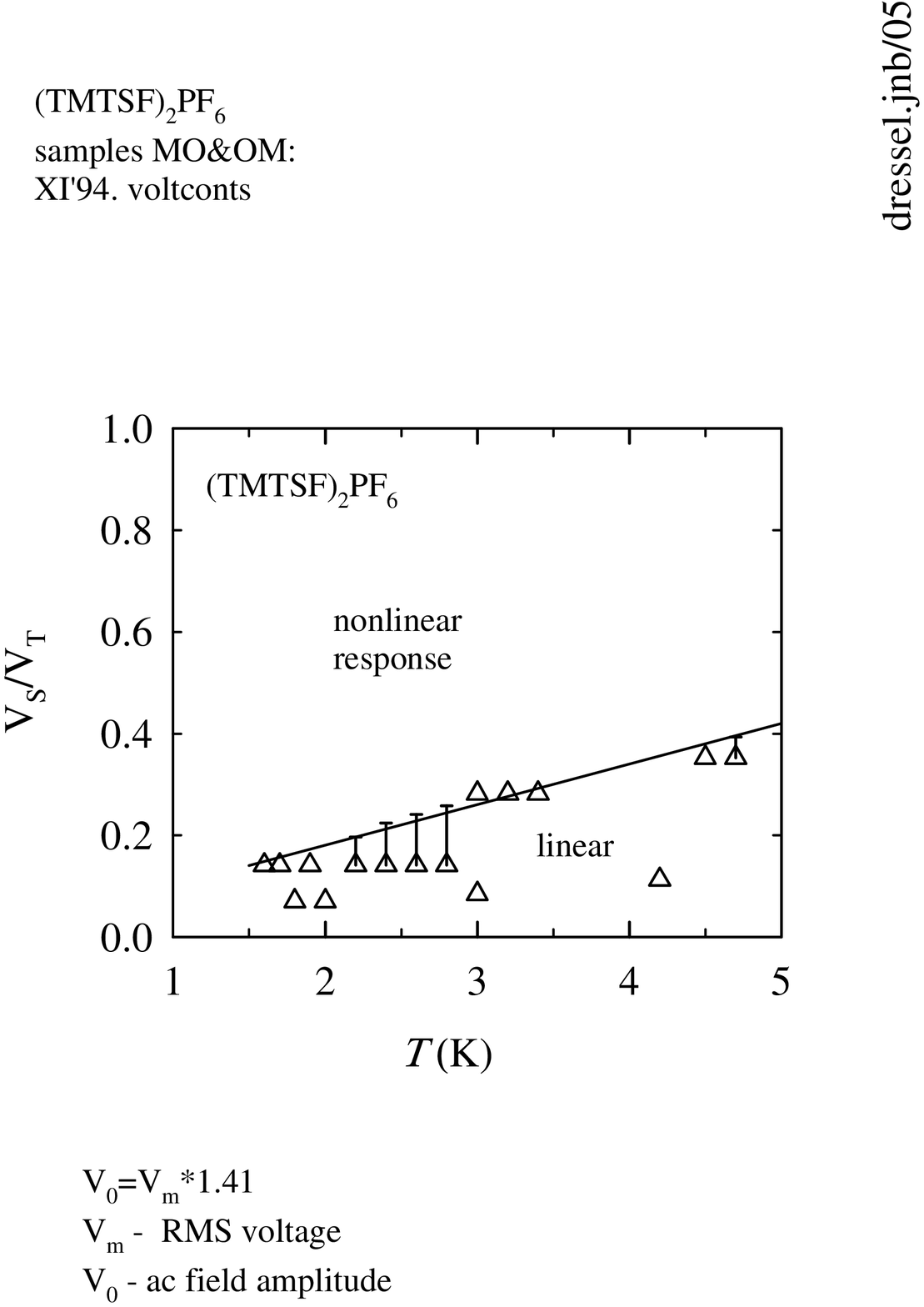}} \caption{ ac-signal
amplitude $V_S$ normalized to the dc threshold $V_T$, vs. temperature.
For $V_S>V_S^{\rm max}$, where $V_S^{\rm max}$ is denoted by the full
line, the sample response starts to deviate from the linear behavior.}
\label{fig:8}
\end{figure}

For a better understanding of the relation between SDW response
and ac signal amplitude, frequency dependent conductivity
measurements along the $a$-direction were performed in the radio
frequency range (20~Hz to 1~MHz) at different temperatures by
changing the amplitude of the ac voltage $V_S$. A deviation from
the linear behavior is observed if the applied ac signal amplitude
$V_S$ increases above a certain threshold voltage $V^{\rm max}_S$
which also strongly depends on temperature (see Fig.~\ref{fig:8}).

\begin{figure}
\centerline{\includegraphics[width=8cm]{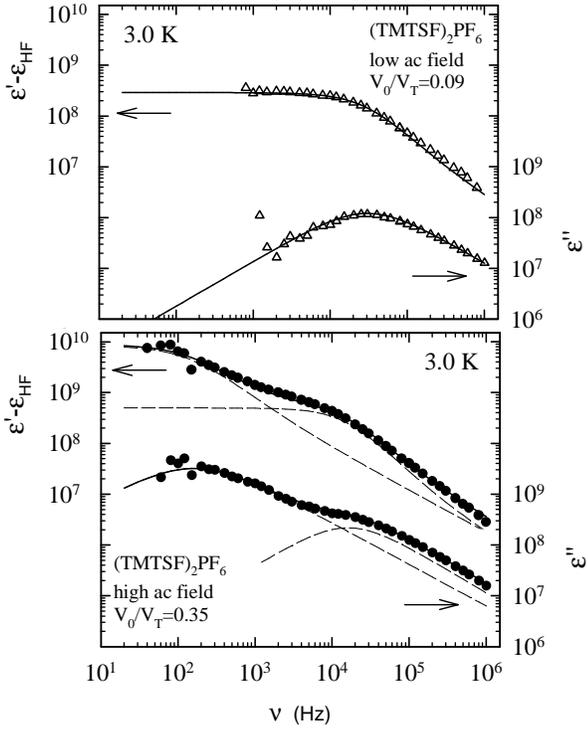}} \caption{ Double
logarithmic plot of the frequency dependence of the real and imaginary
parts of the dielectric function at $T=3$~K for low (upper panel) and
high (lower panel) ac amplitudes. The full lines are fits to the HN
forms; the dashed lines represent the single HN form. } \label{fig:9}
\end{figure}

The upper and lower panels of Fig.~\ref{fig:9} show the real and
imaginary parts of the dielectric constant as a function of frequency
measured at $T=3$~ K for low ($V_S<V_S^{\rm max}$) and high
($V_S>V_S^{\rm max}$) ac amplitudes, respectively. While the observed
dielectric response for very low ac fields was successfully fitted by a
single HN mode of Eq.~(\ref{HN}), similar attempts (dashed lines) to
fit the data measured for higher fields failed. Consequently we tried
to describe the data by the sum of two HN modes and achieved an
excellent fit. The full lines in Fig.~\ref{fig:9} correspond to the
calculated fits to one (upper panel), or sum of the two HN functions
(lower panel). A striking new result is that in addition to the already
reported mode with $\Delta\varepsilon$ of the order of $10^9$, there is
a mode centered at lower frequencies with one order of magnitude larger
strength $\Delta\varepsilon \approx 10^{10}$ which occurs only at
temperatures lower than 4~K.

\begin{figure}
\centerline{\includegraphics[width=8cm]{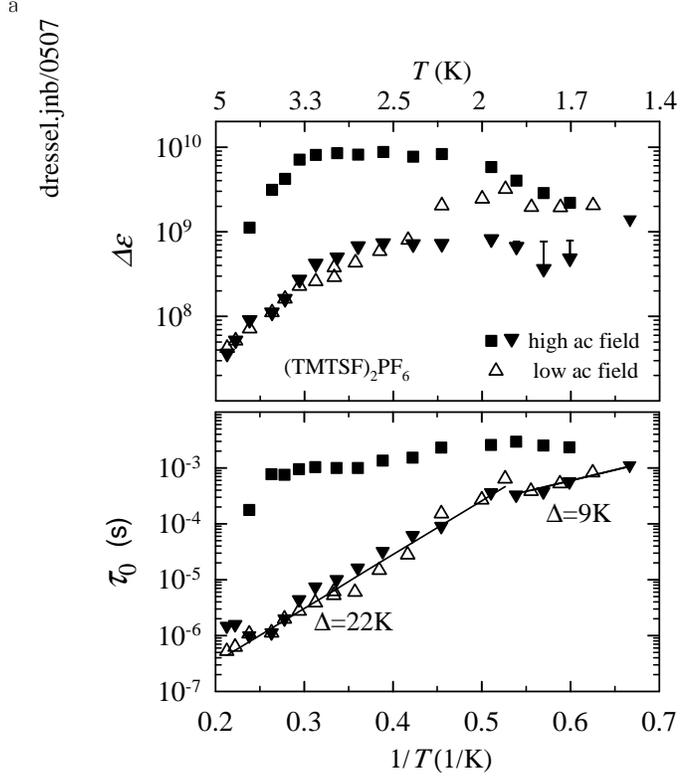}} \caption{
Relaxation strength (upper panel) and mean relaxation time (lower
panel) vs. inverse temperature.  Triangles and squares for HF and the
LF relaxation mode, respectively. Open and full symbols for low and
high ac fields, respectively. } \label{fig:11}
\end{figure}

The dielectric strength $\Delta\varepsilon$ and the mean relaxation
time $\tau_0$ of these two modes are plotted in Fig.~\ref{fig:11} as a
function of inverse temperature. The parameter $1-\alpha$, which
describes the symmetric broadening of the relaxation time distribution
function, is temperatures independent and similar for both modes:
$1-\alpha=0.85$. Triangles (open and full symbols for low and high ac
fields, respectively) stand for the mode already reported previously,
and full squares (for high ac fields) stand for the new mode observed
in this study for the first time. We will refer to these modes as high
frequency (HF) and low frequency (LF) mode, respectively.

There are two effects induced by ac amplitudes $V_S$ larger than
$V_S^{\rm max}$ (see Fig.~\ref{fig:8}): First, the number of
low-frequency relaxations increases, which effectively shifts the SDW
loss peak to lower frequencies. Such an effect was previously observed
for charge density waves and attributed to long-time relaxations from
metastable states far from equilibrium in which the charge density wave
is brought by large ac-signal amplitudes \cite{Cava84}. Second, an
additional mode appears below 4~K, with much longer and
tem\-per\-a\-ture-in\-de\-pend\-ent mean relaxation time $\tau_0
\approx 10^{-3}$ s. At temperatures lower than 2~K, the LF mode
converges to the HF one, so that below 1.7~K only one mode can be
resolved (Fig.~\ref{fig:11}).

\section{Discussion}
\subsection{Anisotropy of the Collective SDW Response}
The significantly reduced values of the activation energy
found in the microwave resistivity right  below the SDW transition
at $T_{\rm SDW}$
for the $a$ and $b^{\prime}$ directions compared to dc data
infer a strong frequency dependent response which
is associated with the collective mode contribution to the
electrical transport.
It can be seen in both orientations, $a$ and $b^{\prime}$, but not in
the $c^*$ direction.

It is worthwhile to consider the entire spectrum of
the SDW response as suggested by G. Gr\"uner and collaborators
\cite{Gruner94b,Donovan94,DresselGruner02}
(Fig.~\ref{fig:10})
Based on an extensive microwave study along the chain direction, it was
proposed \cite{Donovan94} that due to impurity pinning the collective
SDW response in \pf6\ is located around 5~GHz. At first glance, the conductivity below
the energy gap decreases exponentially with decreasing temperature,
except in the range of the pinned mode. As can be seen from
Fig.~\ref{fig:10}, the present microwave experiments are
performed in the range where the collective mode is still very
pronounced, i.e.\ on the high-frequency shoulder of the pinned mode resonance. Hence
the temperature dependent microwave conductivity is caused by two
opposing effects: (i)~the exponential freeze-out of the background
conductivity caused by the uncondensed conduction electrons, and
(ii)~the build-up of the collective contribution. It was suggested that
this mode does not gain much spectral weight as the temperature
decreases, but the width and center frequency changes slightly
\cite{Donovan94}. As discussed in Ref.~\cite{Petukhov04}
the presence of the enhanced microwave conductivity not only along the
chains, but also perpendicular to them implies that the pinned
mode resonance is present in the $b^{\prime}$ direction in a very
similar manner compared to the $a$ axis. Hence the
sliding density wave has to be considered a two-dimensional phenomenon.

These results can be explained by looking at the actual Fermi surface
of \pf6\ which is not strictly one-dimensional but shows a warping in
the direction of $k_b$ (and much less in $k_c$). From NMR experiments
\cite{Takahashi86b} it is known that the SDW corresponds to a
wavevector ${\bf Q} = [0.5a^*,0.24b^*, -0.06c^*]$; which is slightly
incommensurate with the underlying lattice along the $b^*$. Most
important in this context, there is a four-fold periodicity in the real
space not only along the best conductivity $a$-axis, but also
perpendicular to it, i.e.\ along the $b$-axis. The tilt of the nesting
vector is responsible for the similar collective SDW response found in
the microwave experiments along the $a$ and $b^{\prime}$ directions.

\subsection{Field Dependence of the Collective SDW Response}
 It is more difficult to explain the
anomalous microwave response at even lower temperatures ($T<4$~K).
The pinned SDW mode shifts to higher frequencies as the
temperature decreases \cite{Donovan94} and this can even lead to a
decrease of the resistivity at certain frequencies. The
non-monotonic behavior is therefore due to the decreasing single
particle contribution while at the same time the pinned mode
shifts into the relevant frequency window. Increasing the electric
field makes this effect smaller since the pinned mode might remain
at lower frequencies. The spectral weight is likely to be
conserved. The effect can be readily understood in the simple
picture of the washboard potential \cite{Gruner94b} which becomes
anharmonic for larger driving force. In the real material, the
impurity pinning centers are randomly distributed implying a
distribution of metastable states around the equilibrium. This is
reflected in the inhomogenously broadened resonances of the pinned
mode, and in the broad loss peak in the radio-frequency range.
Large ac electric fields, but still smaller than dc threshold for
sliding, can drive the SDW into metastable states with longer
relaxation times far from equilibrium configuration. As a result,
the screened loss peak of the SDW shifts to lower frequencies.
Similar effect can happen to the pinned SDW mode in the case of
high frequencies.  In particular the ac threshold field for
non-linear response is strongly temperature dependent; hence at
low enough temperatures the electric field applied to the sample
inside the cavity could be high enough to exceed $E_T$ for ac
fields. The sample dependence is caused by variations of the
impurity concentration which act as pinning centers.

The deviation from the activated response in the microwave range
sets in pretty abruptly below $T=4$~K, where in the
radio-frequency range the LF mode appears. The relaxation of this
LF mode is governed by very low energy barriers of the order of a
few Kelvin in contrast to the HF process whose relaxation is
governed by the single-particle gap as it is the case for the dc
conductivity. While the over-all behavior of the HF mode can be
ascribed to the screened phason excitations, the spatial
variations of the SDW, which is active in the LF process, should
be restricted to rather narrow localized regions, so that the
associated excitations (like solitons or domain walls) cannot
contribute substantially to the dc conductivity. In addition, the
almost temperature-independent mean relaxation time indicates that
the free-carrier screening, whose influence remains to be visible
in the HF process is insignificant for the LF process. This also
implies that the relaxation entities in the LF process are much
more localized in comparison to the entities responsible for the
HT process \cite{Pinteric01,Pinteric99}. Discommensurations
separating randomly distributed domains of $N=4$ SDW, seem to be
the most plausible candidate. Furthermore, the temperature
behavior of the dielectric strength associated with the LF process
indicates that the mode gains the full spectral weight at about
$T=3.3$~K, and starts to diminish again below 2~K merging
eventually into the HF mode below about 1.6~K.

The observed dielectric and microwave responses seem to be
manifestations of the same phenomenon: i.e.\ the transition from
an incommensurate to a commensurate spin density wave, as
suggested by NMR experiments \cite{Clark93,Takahashi87}. From the
relaxation rate $1/T_1$ of $^1H$-NMR Takahashi {\it et al.}
\cite{Takahashi87} constructed a phase diagram with three SDW
subphases separated at 3.5~K and 1.8~K at ambient pressure. Note
that these temperatures of the second SDW subphase define the
region for which the LF mode is dominant when the ac-amplitude
$V_S$ exceeds $V^{max}_S =0.1-0.3 V_T$.

\section{Conclusions}
The dc and microwave conductivity in the SDW state of \pf6\ single
crystals was measured along all three directions, $a$,
$b^{\prime}$, and $c^*$. Deviations of the microwave response from
the dc activation energy indicate that the pinned mode resonance
is present in the $a$ and $b^{\prime}$ axes response, but not
along the least conducting $c^*$ direction. The behavior can be
explained by the nesting properties of the quasi one-dimensional
conductor.

In addition, the electric-field dependence of the collective
transport was studied in the radiofrequency and microwave range.
The SDW response is non-linear above an ac threshold field because
the SDW is driven into metastable states with longer relaxation
times; this threshold is strongly reduced at temperatures below
4~K. In the non-linear regime, radio-frequency dielectric
measurements have identified, for the first time, two length and
time scales involved in the low-temperature dynamics of the
spin-density-wave state of \pf6. In addition to the well known
screened phason relaxation with Arrhenius-like decay determined by
single-particle gap, there is an additional mode at lower
frequencies with a temperature-indepen\-dent relaxation time found
at temperatures between 4~K and 2~K. We ascribe this mode to
discommensurations, that is, to short-wavelength excitations of
the SDW close to commensurability. With increasing ac field, the
screened phason loss-peak and the pinned-mode resonance shift to
lower frequencies due to increasing contributions of long-time
processes. From our microwave experiments we conclude that this
behavior occurs in the $a$ and $b^{\prime}$ directions. Finally,
below $T=1.7$~K, only one screened phason-like loss-peak can be
detected whose relaxation is governed by low energy barriers of
about half single-particle gap.

\section*{Acknowledgment}
We thank Gabriele Untereiner for the crystal growth and sample
preparation. The work was supported by the Deut\-sche
Forschungs\-ge\-mein\-schaft (DFG) and Croatian Ministry of
Science.


\begin{thebibliography}{99}
\bibitem{Jerome82}
D. J{\'e}rome and H.J. Schulz, {Adv. Phys.} {\bf 31},  299  (1982).
\bibitem{Gruner94a}G. Gr\"uner, Rev. Mod. Phys. {\bf 66}, 1 (1994).
\bibitem{Gruner94b}
G. Gr{\"u}ner, {\em Density Waves in Solids} (Addison-Wesley, Reading,
1994).
\bibitem{Takahashi86b}
T. Takahashi, Y. Maniwa, H. Kawamura, and G. Saito, J. Phys. Soc. Jpn.
{\bf 55},  1364  (1986).
\bibitem{Degiorgi96}
L. Degiorgi, M. Dressel, A. Schwartz, B. Alavi,
and G. Gr\"{u}ner, Phys. Rev. Lett. {\bf 76}, 3838 (1996);
M. Dressel, L. Degiorgi, J. Brinkmann, A. Schwartz, and G. Gr\"{u}ner,
Physica B {\bf 230-232}, 1008 (1997);
V. Vescoli, L. Degiorgi, M. Dressel, A. Schwartz, W. Henderson,
B. Alavi, G. Gr\"{u}ner, J. Brinkmann, and A. Virosztek,
Phys. Rev. B {\bf 60}, 8019 (1999).
\bibitem{Mihaly97}G. Mih{\'a}ly, A. Virosztek, and G. Gr{\"u}ner,
Phys. Rev. B {\bf 55}, 13456 (1997).
\bibitem{Donovan94}
S.~Donovan, M.~Dressel, Y.~Kim, L.~Degiorgi, G.~Gr\"{u}ner, and W.~Wonneberger,
Phys.~Rev. B {\bf 49}, 3363 (1994).
\bibitem{Tomic89}
S.Tomi{\'c}, J.R.Cooper, D.J{\'e}rome and K.Bechgaard,
Phys.Rev.Lett.{\bf 62}, 462 (1989);
W. Kang, S. Tomi{\'c}, and D. J{\'e}rome, Phys. Rev. B
{\bf 43}, 1264 (1991); S. Tomi{\'c}, J.R. Cooper, W. Kang, D. J{\'e}rome, and
K. Maki, J. Phys. I (France) {\bf 1}, 1603 (1991).
\bibitem{Mihaly91}
G. Mih{\'{a}}ly, Y. Kim, and G. Gr{\"{u}}ner, Phys. Rev. Lett. {\bf 67},
2713  (1991); {\em ibid.} {\bf 66},  2806  (1991);
O. Tr{\ae}tteberg, G. Kriza, and G. Mih{\'{a}}ly, Phys. Rev. B {\bf 45},
8795  (1992); F. Z{\'a}mborszky, G. Szeghy, G. Abdussalam, G. Mih{\'a}ly,
Phys. Rev. B {\bf 60}, 4414 (1999).
\bibitem{Nomura89}K. Nomura, T. Shimizu, K. Ichimura, T. Sambongi, M.
Tokumoto, H. Anzai, and N. Kigoshita, Solid State Commun. {\bf 72},
1123 (1989); M.Basleti\'{c}, N.Bi\v{s}kup, B.Korin-Hamzi\'{c}, A.Hamzi\'{c} and S.Tomi\'{c},
Fizika A {\bf 8}, 293-310 (2000).
\bibitem{Kriza91}G. Kriza, G. Quirion, O. Tr{\ae}tteberg, W. Kang, and D.
J\'erome, Phys. Rev. Lett. {\bf 66}, 1922 (1991).
\bibitem{Clark93}
W.G. Clark, M.E. Hanson, W.H. Wong, and B. Alavi, {J. Phys. IV. (France)}
 {\bf 3},  C2 235  (1993); Physica B {\bf 194-196},   285  (1994);
W.H. Wong, M.E. Hanson, B. Alavi, W.~G. Clark, and W.~A. Hines,
{Phys. Rev. Lett.} {\bf 70},  1882  (1993);
W.H. Wong, M.E. Hanson, W.G. Clark, B. Alavi, and G. Gr{\"u}ner, Phys.
Rev. Lett. {\bf 72},  2640  (1994).
\bibitem{Takahashi87}T. Takahashi, Y. Maniwa, H. Kawamura, K. Murata, and G.
Saito, Synth. Met. {\bf 19}, 225 (1987); T. Takahashi, T. Harada, Y. Kobayashi,
K. Kanoda, K. Suzuki, K. Murata, and G. Saito, Synth. Met. {\bf 41}, 3985 (1991).
\bibitem{Odin94}
J.~Odin, J.C.~Lasjaunias, K.~Biljakovi\'{c}, P.~Monceau and
K.~Bechgaard, Solid State Comm.\ {\bf 91}, 523 (1994).
\bibitem{Basletic02}
M.~Basletic, B.~Korin-Hamzic and K.~Maki, Phys.\ Rev.\ B {\bf 65}, 235117 (2002).
\bibitem{Dressel05}M. Dressel, K. Petukhov, B. Salameh, P. Zornoza, and T.
Giamarchi, Phys. Rev. B (in press 2005); cond-mat/0409322.
\bibitem{Pinteric01}
M.Pinteri\'{c}, T.Vuleti\'{c}, S.Tomi\'{c} and J.U.von Sch\"{u}tz, Eur.\ Phys.\ J.\ B {\bf 22}, 335 (2001).
\bibitem{Klein93}O.~Klein, S. Donovan, M. Dressel, and G. Gr{\"u}ner,
{Int.~J. Infrared and Millimeter Waves}, {\bf 14}, 2423 (1993);
S.~Donovan, M. Dressel, O. Klein, K. Holczer, and G. Gr{\"u}ner,
{Int.~J. Infrared and Millimeter Waves}, {\bf 14}, 2459 (1993);
M.~Dressel, O. Klein, S. Donovan, and G. Gr{\"u}ner, {Int.~J.
Infrared and Millimeter Waves}, {\bf 14}, 2489 (1993).
\bibitem{DresselGruner02}
M.~Dressel and G.~Gr\"{u}ner, {\it Electrodynamics of
Solids}  (Cambridge University Press, Cambridge, 2002).
\bibitem{Petukhov04}
K. Petukhov and M. Dressel, Phys. Rev. B (in press 2005);
cond-mat/0408382.
\bibitem{remark2}Note, the microwave data displayed in Fig.~\protect\ref{fig:2},
\protect\ref{fig:3}, and \protect\ref{fig:5} are taken
at different samples; their absolute values of resistivity vary. For
microwave experiments it is also known that the
absolute values also contain a large uncertainty due to the geometry
(depolarization factor) \protect\cite{Klein93}.
\bibitem{Walsh80}W.M.~Walsh Jr., F.~Wudl, G.A.~Thomas,
D.~Nalewajek, J.J.~Hauser, P.A.~Lee, and T.~Poehler,
Phys.~Rev.~Lett. {\bf 45}, 829 (1980).
\bibitem{Janossy83}A.~J\`{a}nossy, M.~Hardiman, and G.~Gr\"{u}ner,
Solid State Commun. {\bf 46}, 21 (1983).
\bibitem{Buravov85}L.I.~Buravov, V.N.~Laukhin, and A.G.~Khomenko,
Sov.~Phys.~JETP {\bf 61}, 1292 (1985).
\bibitem{Javadi85}H.H.S.~Javadi, S.~Sridhar, G.~Gr\"{u}ner,
L.~Chiang, and F.~Wudl,  Phys.~Rev.~Lett. {\bf 55}, 1216
(1985).
\bibitem{remark1}The only exception is the 3~GHz measurement, which
was performed in a split ring resonator and not a enclosed cavity.
\bibitem {Cava84}
R.J.Cava, R.M. Flemming, R.G. Dunn, E.A. Rietman, and L.F. Schneemeyer,
Phys.\ Rev.\ B {\bf 30}, 7290 (1984).
\bibitem{Pinteric99}
M.~Pinteri{\'c}, M.~Miljak, N.~Bi{\v s}kup, O.~Milat, I.~Aviani,
S.~Tomi{\'c}, D.~Schweitzer, W.~Strunz and I.~Heinen, Eur.\ Phys.\
J.\ B \textbf{11}, 217 (1999).
\end{thebibliography}
\end{document}